\documentclass[11pt]{article}
\usepackage{blois,epsfig}

\def\be{\begin{equation}}
\def\ee{\end{equation}}
\def\bea{\begin{eqnarray}}
\def\eea{\end{eqnarray}}

\begin{document}
\vspace*{4cm}
\title{SUPERSYMMETRY IN Z' DECAYS}

\author{Gennaro Corcella}
\footnote{
Talk given at the 
24th Rencontres de Blois, Ch\^ateau de Blois, France, May 27 - June 1,
2012.}

\address{INFN, Laboratori Nazionali di Frascati,\\
Via E. Fermi 40, I-00044, Frascati (RM), Italy}

\maketitle\abstracts{I study the phenomenology 
of new heavy neutral gauge bosons $Z'$, predicted by
GUT-driven U(1)$'$ gauge groups and by the
Sequential Standard Model.  BSM decays into supersymmetric
final states are accounted for, besides the SM modes
so far investigated. I give an estimate of the 
number of supersymmetric events in $Z'$ decays possibly expected 
at the LHC, as well as of the product of 
the $Z'$ cross section times the branching fraction into 
electron and muon pairs.}

Heavy neutral gauge bosons $Z'$ are predicted by
gauge groups U(1)$'$, which arise in the framework of
Grand Unification Theories (GUTs) \cite{rizzo,langacker}. 
They are also present
in the Sequential Standard Model (SSM), the simplest extension of the
Standard Model (SM), as it just contains an extra heavy neutral boson 
with the same couplings
to fermions and bosons as the $Z$.
Both Tevatron \cite{tevatron} and LHC \cite{lhc} experiments 
have searched for high-mass electron or muon 
pairs, which are the typical
signature of $Z'$ decays.
As for the LHC,  CMS and ATLAS excluded a 
U(1)$'$-based $Z'$
with mass below 2.32 TeV and 2.21 TeV, respectively, and a  
SSM $Z'$ below
1.49-1.69 TeV (CMS) and 1.77-1.96 TeV (ATLAS).
More recently, $\tau^+\tau^-$ pairs were also considered to 
exclude a $Z'$ boson with mass below 1.1-1.4 TeV \cite{tau}.

These analyses assume, however, that the $Z'$ has only
SM decay modes. 
Supersymmetric contributions to $Z'$ decays may possibly
decrease the SM branching ratios 
and therefore have an impact on the exclusion limits on the 
$Z'$ mass. In this talk I will highlight recent 
results \cite{corge} on supersymmetry
signals in $Z'$ decays, referring
to the Minimal Supersymmetric Standard Model (MSSM)
and updating pioneering work on this topic 
\cite{gherghetta}.

Other studies lately 
carried out \cite{baum,chang} will be improved by 
consistently including 
the so-called D-term correction, due to the extra U(1)$'$ group, 
to the sfermion masses. In fact, the D-term can be large
and negative, in such a way to drive the sfermion squared masses
to become negative and thus unphysical.
Furthermore, 
the supersymmetric particle masses 
will not be treated as free parameters, but 
will be obtained by diagonalizing the
corresponding mass matrices.

The U(1)$'$ models originate from the breaking of a rank-6
GUT group E$_6$ according to ${\rm E}_6\to {\rm SO}(10)\times
{\rm U}(1)'_\psi$, followed by ${\rm SO} (10)\to {\rm SU}(5)
\times {\rm U}(1)'_\chi$.
The heavy neutral bosons associated with ${\rm U}(1)'_\psi$ and 
${\rm U}(1)'_\chi$ are then 
named $Z'_\psi$ and $Z'_\chi$,
whereas a generic $Z'$ boson is a combination of $Z'_\psi$ and $Z'_\chi$,
with a mixing angle $\theta$:
\begin{equation}
Z'(\theta)=Z'_\psi\cos\theta-Z'_\chi\sin\theta.
\end{equation}
The $Z'$ bosons and the $\theta$ values which will be
investigated are listed in Table~\ref{tabmod}.
The $Z'_\eta$ model comes from the direct breaking of the GUT group in
the SM, i.e. ${\rm E}_6\to {\rm SM} 
\times {\rm U}(1)'_\eta$;
the $Z'_{\rm S}$ is present in the secluded model, wherein the SM is extended by means of
a singlet field $S$; the $Z'_{\rm N}$ is instead
equivalent to the $Z'_\chi$ 
model,
but with the `unconventional' assignment of SM, MSSM 
and exotic fields 
in the SU(5) representations \cite{nardi}.
\begin{table}
  \caption{$Z'$ bosons in the U(1)$'$ models along with the 
mixing angles.  \label{tabmod}}\vspace{0.2cm}
\begin{center}
  \begin{tabular}{|c|c|}
\hline
    Model & $\theta$ \\
\hline
$Z'_\psi$ & 0\\
\hline
$Z'_\chi$ & $-\pi/2$\\
\hline
$Z'_\eta$ & $\arccos\sqrt{5/8}$\\ 
\hline
$Z'_{\rm S}$ & $\arctan(\sqrt{15}/9)-\pi/2$\\
\hline
$Z'_{\rm I}$ & $\arccos\sqrt{5/8}-\pi/2$\\
\hline
$Z'_{\rm N}$ & $\arctan\sqrt{15}-\pi/2$\\
\hline
  \end{tabular}
\end{center}
\end{table}
When extending the
MSSM with a $Z'$ boson, one has two extra neutralinos, for a total
of six neutralinos, and an extra neutral scalar Higgs.
These new Higgs and neutralinos must be included
when studying $Z'$ decays within the MSSM and U(1)$'$ framework;
however, they are typically quite heavy and therefore their 
contribution to the
$Z'$ width is rather small \cite{corge}.

In the extended MSSM, besides the SM modes, one has to
consider $Z'$ decays into slepton, squark, chargino, neutralino 
and Higgs pairs,
as well as final 
states with Higgs bosons associated with a $W$ or a $Z$.
Decays
yielding  charged leptons are the golden channels for the experimental
searches and deserve special attention.
$Z'$ decays into charged sleptons 
$Z'\to\tilde\ell^+\tilde\ell^-$, with the sleptons decaying 
according to $\tilde\ell^\pm\to \ell^\pm\tilde\chi^0$, 
$\tilde\chi^0$ being a neutralino, or chargino modes like 
$Z'\to\tilde\chi_2^+\tilde\chi_2^-$, 
followed by $\tilde\chi_2^\pm\to \ell^\pm\tilde\chi_1^0$,
lead to a final state with two charged leptons and missing energy.
Four leptons and missing energy can be given by the decay chain
$Z'\to\tilde\chi_2^0\tilde\chi_2^0$, with
subsequent $\tilde\chi_2^0\to \ell^\pm\tilde\ell^\mp$ and 
$\tilde\ell^\pm\to\ell^\pm\tilde\chi_1^0$.
Finally, sneutrino-pair production, i.e.
$Z'\to\tilde\nu_2\tilde\nu_2^*$, followed by  $\tilde\nu_2\to  \tilde\chi^0_2\nu$
and $\tilde\chi_2^0\to \ell^+\ell^-\tilde\chi^0_1$, 
with an intermediate charged slepton, yields four
charged leptons and missing energy (neutrinos and neutralinos).

A point in the parameter space can be chosen to
compute the masses of the supersymmetric
particles 
and study the dependence of the 
branching ratios on the MSSM and U(1)$'$ parameters.
Particular care has been taken about the relevance of 
$Z'$ and slepton masses in the decay rates
in the channels eventually leading to leptons and 
missing energy. 
Hereafter, this study will be carried out setting the 
U(1)$'$/MSSM parameters to the 
following `Reference Point' \cite{corge}:
\begin{eqnarray}
&\ &
\mu =200\ ,\ \tan\beta=20\ ,\ A_q=A_\ell=500~{\rm GeV}\ ,\nonumber \\
&\ &m^0_{\tilde q}=5~{\rm TeV}\ ,\ 
 M_1=150~{\rm GeV}\ ,\ M_2=300~{\rm GeV}\ ,\ M^\prime=1~{\rm TeV}.
\label{refpoint}
\end{eqnarray}
In Eq.~(\ref{refpoint}), $\mu$ is the well-known parameter
contained in the 
Higgs superpotential, $\tan\beta=v_2/v_1$ is the ratio of the vacuum
expectation values of the two MSSM Higgs doublets, 
$A_q$ and $A_\ell$ are the couplings of the Higgs with quarks
and leptons, respectively.
Furthermore, $m^0_{\tilde q}$ is the squark mass, assumed
to be the same for all flavours at the $Z'$ scale, before the
addition of the D-term, $M_1$, $M_2$ and $M'$ are the soft masses of
the gauginos $\tilde B$, $\tilde W_3$ and $\tilde B '$.
The U(1)$'$ coupling $g'$, as occurs in GUTs,
is proportional to the standard U(1) coupling constant
$g_1$ via $g'=\sqrt{5/3}\ g_1$.
In the Sequential Standard Model, 
the coupling of the $Z'_{\rm SSM}$ to the fermions
is the same as in the SM, i.e. $g_{\rm{SSM}}=g_2/(2\cos\theta_W)$,
where $g_2$ is the SU(2) coupling and $\theta_W$ the Weinberg angle.
As for $m^0_{\tilde\ell}$, the slepton mass before the D-term
contribution, fixed to the same value for selectrons, 
smuons, staus and sneutrinos, 
it can be determined in such a way that,
for a given $Z'$ mass in the range
1 TeV$<m_{Z'}<$ 5 TeV, the rate of the
decay $Z'\to \tilde\ell^+\tilde\ell^-$ is maximized 
\cite{corge}.
With the parametrization (\ref{refpoint}) and this choice,
the total branching ratio into BSM channels can be up to about 60\%
($Z'_{\rm{SSM}}$), 40\% ($Z'_\psi$), 30\% ($Z'_\eta$ and 
$Z'_{\rm N}$), 20\% ($Z'_{\rm S}$) and 15\% ($Z'_{\rm I}$).
The highest branching ratios are those into chargino and
neutralino pairs, which can account up to 20\% and 30\% of the total
$Z'$ width, respectively. Direct decays into charged-slepton
pairs have branching fractions of the order of few percent
\cite{corge}.

In Table~\ref{number} I present the expected number of
events with supersymmetric cascades ($N_{\rm casc}$), 
i.e. production of
neutralinos, charginos or sleptons, 
and the charged-slepton rates ($N_{\rm slep}$), at the LHC
for an integrated luminosity ${\cal L}=20~{\rm fb}^{-1}$
and a centre-of-mass energy $\sqrt{s}=8$~TeV. 
All parameters are fixed to
the Reference Point (\ref{refpoint}); 
the $Z'$ mass is set to either 1.5 or
to 2 TeV, whereas $m^0_{\tilde\ell}$ is fixed to the value which 
enhances the $Z'\to\tilde\ell^+\tilde\ell^-$ decay
rate \cite{corge}. 
The numbers in Table~\ref{number} are obtained in the
narrow-width approximation, with the $pp\to Z'$ cross section
computed at leading order using the CTEQ6L parton 
distribution functions \cite{cteq}. The $Z'_\chi$ model
is not taken into account, since, using the parametrization
(\ref{refpoint}), it does not lead to a physical sfermion
spectrum.
One finds that the cascade events can be  
${\cal O}(10^3)$ and the charged sleptons
up to a few dozens: the highest rate of production
of supersymmetric particles occurs in the SSM,
but even the U(1)$'$ models yield meaningful sparticle production.
\begin{table}
  \caption{Number of supersymmetric cascade events
 and charged sleptons at the LHC
for an integrated luminosity of 20 fb$^{-1}$ and a 
centre-of-mass energy
of 8 TeV. The $Z'$ mass is quoted in TeV.}\vspace{0.2cm}
\begin{center}  
\begin{tabular}{|c|c|c|c|c|c|}
\hline
Model & $m_{Z'}$ & N$_{\rm casc}$ & N$_{\rm slep}$ \\ 
\hline
$Z'_\eta$ & 1.5  & 523  & -- \\
\hline
$Z'_\eta$ & 2 &  55 & -- \\
\hline
$Z'_\psi$ & 1.5 &  599 & 36 \\
\hline
$Z'_\psi$ & 2 & 73 & 4 \\
\hline
$Z'_{\rm N}$ & 1.5 &  400 & 17 \\
\hline
$Z'_{\rm N}$ & 2 &  70 & 3 \\
\hline
$Z'_{\rm I}$ & 1.5  &  317 & -- \\
\hline
$Z'_{\rm I}$ & 2  & 50 & -- \\
\hline
$Z'_{\rm S}$ & 1.5  & 30 & -- \\
\hline
$Z'_{\rm S}$ & 2  & 46  & -- \\
\hline
$Z'_{\rm SSM}$ & 1.5  &  2968  & 95 \\
\hline
$Z'_{\rm SSM}$ & 2  & 462 & 14 \\
\hline
  \end{tabular}
\end{center}
  \label{number}
\end{table}

\begin{figure}[ht]
\centerline{\resizebox{0.5\textwidth}{!}{\includegraphics{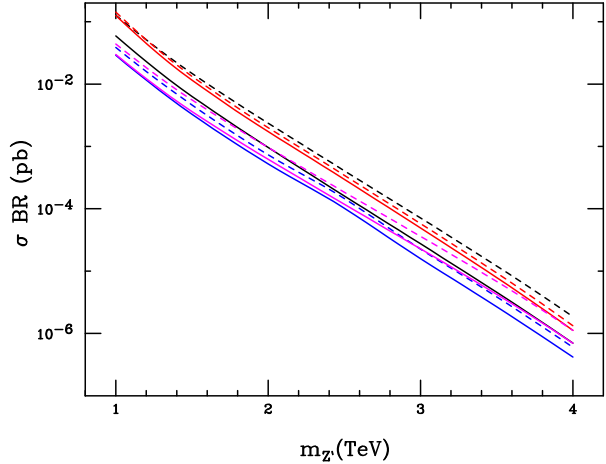}}}
\caption{Product of the cross section ($\sigma$) and the branching ratio (BR)
into $e^+e^-$ and $\mu^+\mu^-$ pairs for 
$Z'$ production in $pp$ collisions at $\sqrt{s}=8$~TeV,
according to the models $Z'_{SSM}$ (black online), $Z'_\eta$ (blue),
$Z'_{\rm I}$ (red) and 
$Z'_\psi$ (magenta). The solid lines account for BSM decay modes, the dashes
just rely on SM channels.}
\label{sbr}
\end{figure}

Before concluding, since the experimental analyses search for high-mass
dielectron and dimuon pairs, I present the results 
in terms of the product of the $Z'$ production
cross section ($\sigma$) at the LHC ($\sqrt{s}=8$~TeV)
times the branching ratio (BR) into $e^+e^-$ and $\mu^+\mu^-$
pairs, with and without accounting for 
supersymmetric modes.
In the Sequential Standard Model, 
assuming only SM decays, it is BR~$\simeq 6.8\%$, like
the SM $Z$ boson: the addition of supersymmetric modes decreases
the lepton rates and may therefore have an impact on the limits on
the $Z'$ mass.
In fact, the experimental exclusion limits are obtained by comparing
data and theoretical predictions for $\sigma $~BR.
Fig.~\ref{sbr} shows this product 
varying the $Z'$ mass in the range 1 TeV$<m_{Z'}<$ 4 TeV,
with only SM decays (dashes) and accounting for possible
supersymmetric contributions (solid lines). In the BSM case,
the parameters are fixed to the Reference Point (\ref{refpoint}),
with the slepton mass $m^0_{\tilde\ell}$ set as explained above
\cite{corge}.
One can thus note that, when including the 
BSM decay modes, the suppression of $\sigma$~BR is about 60\% for 
the $Z'_{\rm SSM}$, 40\% for the $Z'_\psi$ model, 30\% for the $Z'_\eta$ and
13\% for the $Z'_{\rm I}$.
Such results point out a remarkable impact of the inclusion of the
supersymmetric contributions to $Z'$ decays and 
that a novel analysis, taking into account
such modes, may be worthwhile to be pursued.

In summary, I briefly discussed possible supersymmetric 
contributions to
$Z'$ decays, within the Minimal Supersymmetric Standard Model,
extended by means of an extra GUT-inspired
U(1)$'$ group, as well as in the Sequential Standard Model.
The next step of 
this investigation will consist in the implementation of 
$Z'$ production and decay modes 
in the framework of a Monte Carlo event generator, in such a way
to account for parton showers, hadronization, underlying event,
finite-width effects and possibly detector simulations.
One should then be able to compare the supersymmetry signals in
$Z'$ decays with the Standard Model backgrounds.
In this way, a conclusive statement on the
LHC reach for $Z'$ bosons within supersymmetry can ultimately
be drawn.
This work is in progress.

\section*{Acknowledgments}
The presented results have been obtained in collaboration with
Simonetta Gentile. I also acknowledge Daniel Froidevaux, who suggested
the study of the product of the cross section times the 
dilepton branching ratio.

\end{document}